\newcommand{\rem}[1]{}
\newcommand{\beq}{\begin{equation}}
\newcommand{\eeq}{\end{equation}}
\newcommand{\beqa}{\begin{eqnarray}}
\newcommand{\eeqa}{\end{eqnarray}}
\newcommand{\ba}{\begin{array}}
\newcommand{\ea}{\end{array}}

\documentclass[pra,twocolumn,showpacs,preprintnumbers,
amsmath,amssymb,floatfix]{revtex4}
\usepackage{hyperref}
\usepackage{epsfig}

\begin{document}

\title{Enhancement of four reflection shifts by a three-layer 
surface plasmon resonance} 

\author{Luca Salasnich}
\affiliation{Dipartimento di Fisica e Astronomia ``Galileo Galilei'' and 
CNISM, Universit\`a di Padova, Via Marzolo 8, 35131 Padova, Italy}

\date{\today}

\begin{abstract}
We investigate the effect of a surface 
plasmon resonance on Goos-Hanchen and Imbert-Fedorov spatial and 
angular shifts in the reflection of a light beam by considering 
a three-layer system made of glass, gold and air. 
We calculate these spatial and angular shifts as a function 
of the incidence angle showing that they are strongly enhanced 
in correspondence of the resonant angle.  
In particular, we find giant spatial and angular Goos-Hanchen 
shits for the p-wave light close to the plasmon resonance. 
We also predict a similar, but less pronounced, resonant effect 
on spatial and angular Imbert-Fedorov shifts 
for both s-wave and p-wave light. 

\end{abstract}


\maketitle

It is now established 
\cite{goos,artmann,imbert,fedorov,exp-angular1,exp-angular2,established} 
that there are four 
shifts that can happen when light is reflected. 
These are the two longitudinal shifts (spatial and angular GH shifts) 
and the two transverse shifts (spatial and angular IF shifts). 
Recently, the predicted GH spatial shift \cite{wild} of a light beam on 
a metallic mirror has been measured \cite{merano}, 
showing a good agreement with the theory. 
Moreover, giant GH spatial shifts have been observed 
\cite{zhang} with a three-layer system in the Kretschmann-Raether 
configuration at the metal-air interface when the surface plasmon resonance of 
the metal is excited. Finally, 
the detection of both GH and IF spatial shifts has been reported \cite{pillon} 
for a light beam on a three-layer system totally reflected 
on the external interface of a dielectric thin film deposited 
on a high-index substrate. 

Motivated by these experimental achievements 
in this paper we analyze theoretically the effect of a surface 
plasmon resonance not only on GH and IF spatial shifts but also on 
GH and IF angular shifts. In particular, we investigate the three-layer 
system glass-gold-air in the Kretschmann-Raether 
configuration \cite{raether,guo}. To calculate spatial and angular 
shifts we use the recently established theory \cite{aiello}, 
which shows that both spatial and angular GH and IF shifts can be derived 
in terms of the complex reflection coefficient. 
We consider the three-layer system glass-gold-air. 
This is the familiar Kretschmann-Raether 
configuration \cite{raether}:  
the light beam comes from the prism of glass and 
reflects with incident angle $\theta$ 
at the interface with air, where there is 
a thin film of gold with thickness $d$. In our anaysis 
the prism of glass has relative permittivity   
$\epsilon_0=2.19$;  the thin film of gold has complex 
relative permittivity: $\epsilon_1=-29.02+2.03 i$ 
for a wavelength  $\lambda=830$ nm \cite{merano}; 
the air has relative permittivity $\epsilon_2=1$. 

The s-wave and p-wave reflection coefficients $r_s$ and $r_p$ 
of this three-layer system 
can be written in terms of generalized Fresnell equations \cite{raether}
\beqa 
r_s &=& {r_s^{01} + r_s^{12} e^{2 i \delta} 
\over 1 + r_s^{01} r_s^{12} e^{2 i \delta} } \; , 
\label{gf1}
\\
r_p &=& {r_p^{12} + r_p^{12} e^{2 i \delta} 
\over 1 + r_p^{01} r_p^{12} e^{2 i \delta} } \; , 
\label{gf2}
\eeqa
where 
\beqa 
r_s^{01}={k_{z0}-k_{z1} \over k_{z0}+k_{z1}} \; , 
\\
r_s^{12}={k_{z1}-k_{z2} \over k_{z1}+k_{z2}} \; , 
\\
r_p^{01}={\epsilon_1 k_{z0}-\epsilon_0 k_{z1} 
\over \epsilon_1 k_{z0}+\epsilon_0 k_{z1}} \; , 
\\
r_p^{12}={\epsilon_2 k_{z1} -\epsilon_1 k_{z2} \over 
\epsilon_2 k_{z1} + \epsilon_1 k_{z2}} \;  
\eeqa
are the reflection coefficient at the $01$ and $12$ interfaces, 
and 
\beqa
k_{z0} &=& {2\pi\over \lambda} 
\sqrt{\epsilon_0} \cos(\theta) \; , 
\\
k_{z1} &=& {2\pi\over \lambda} 
\sqrt{\epsilon_1-\epsilon_0 \sin^2(\theta)} = {\delta\over d} \; , 
\\
k_{z2} &=& {2\pi\over \lambda} 
\sqrt{\epsilon_2-\epsilon_0 \sin^2(\theta)} \;   
\eeqa 
are the $z$ components of the wavevectors of the light. 
Notice that $k_{z1}$ gives the ratio between the 
complex phase parameter $\delta$ 
which appears in Eqs. (\ref{gf1}) and (\ref{gf2}) 
and the thickness $d$ of the gold film. 

\begin{figure}
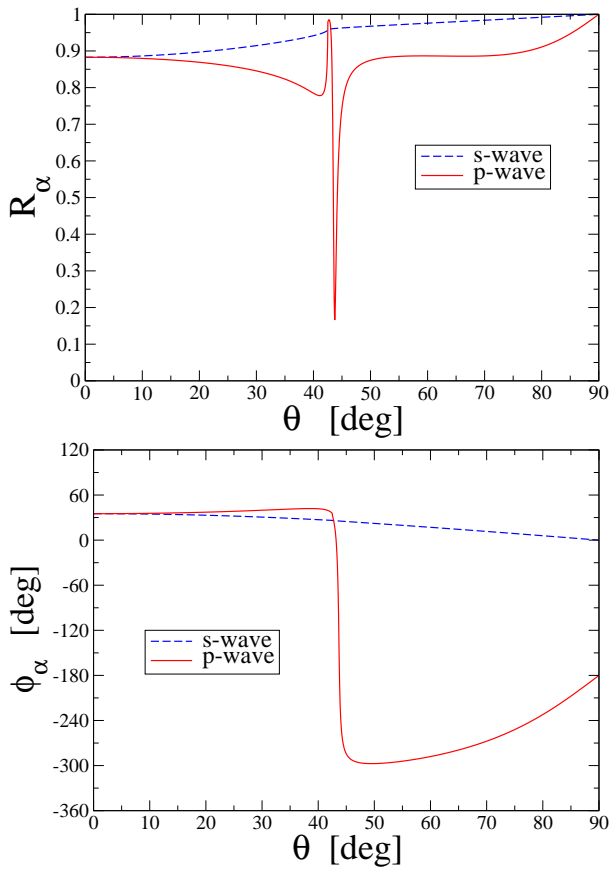

\centerline{\epsfig{file=reflect.eps,width=8cm,clip=}}
\centerline{\epsfig{file=phase.eps,width=8cm,clip=}}
\caption{(Color online). Upper panel: 
Reflectivity $R_{\alpha}$ as a function of the 
incident angle $\theta$ for s-wave ($\alpha=s$, dashed line) and p-wave 
($\alpha=p$, solid line) monochromatic light. 
Lower panel: Phase $\phi_{\alpha}$ of the reflection coefficient 
$r_{\alpha}=R_{\alpha}^{1/2} \ e^{i\phi_{\alpha}}$ as a function of 
the incident angle $\theta$ 
for s-wave ($\alpha=s$, dashed line) and p-wave 
($\alpha=p$, solid line) monochromatic light. 
For both panels: Three-layer system with a gold film of thickness $d=30$ nm 
as glass-air interface and light with wavelength $\lambda=830$ nm.}
\label{fig1}
\end{figure}

\begin{figure}
\centerline{\epsfig{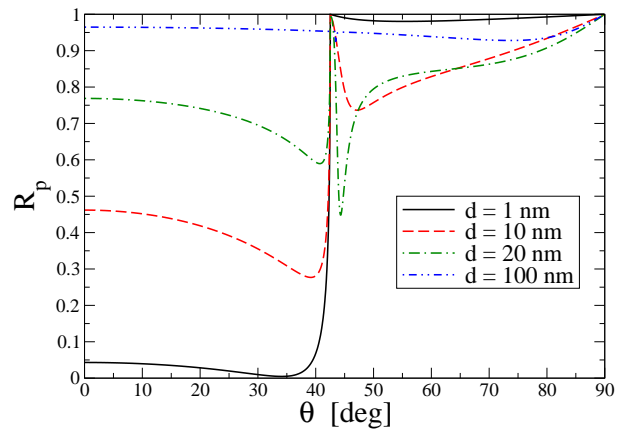}}
\caption{(Color online). P-wave reflectivity $R_{p}$ as a function of the 
incident angle $\theta$ for monochromatic light: 
results for four values of the thickness $d$ of the gold film. 
Three-layer system with a gold film 
as glass-air interface and light with wavelength $\lambda=830$ nm.}
\label{fig2}
\end{figure}

It is important to stress that, in general, both s-wave and p-wave 
reflection coefficients $r_s$ and $r_p$ of Eqs. (\ref{gf1}) and (\ref{gf2}) 
are complex numbers. Thus one usually writes them as 
\beq
r_{\alpha} = R_{\alpha}^{1/2} \ e^{i \phi_{\alpha}} \; ,
\label{hope}
\eeq
where $R_{\alpha}=|r_{\alpha}|^2$ is the 
$\alpha$-wave ($\alpha=s,p$) reflectivity 
($0\leq R_{\alpha} \leq 1$) and $\phi_{\alpha}$ is the 
corresponding reflection phase. In the upper panel of 
Fig. \ref{fig1} we report the reflectivity $R_{\alpha}$, 
obtained directly from Eqs. (\ref{gf1}) and (\ref{gf2}),  
as a function of the incident angle $\theta$ 
for s-wave (dashed line) and p-wave (solid line) 
monochromatic light of wavelength $\lambda=830$ nm and choosing 
the thickness $d=30$ nm for the gold film. The panel shows the 
well-known surface plasmon resonance at $\theta = \theta_R \simeq 43.7^o$ 
\cite{raether}. This strong reduction of the p-wave reflectivity $R_p$ around 
the resonant angle $\theta_R$ is due to the formation 
of an evanescent wave which propagates along the interface 
between gold and air \cite{raether}. Both the reduction of 
reflectivity and the position of resonant angle $\theta_R$ depend 
on the thickness $d$ of the metal. Our choice $d=30$ nm gives a 
p-wave reflectivity $R_p$ close to zero at the resonant angle $\theta_R$. 
The figure clearly shows that the s-wave reflectivity is a monotonic 
increasing function of $\theta$. Instead the p-wave reflectivity 
has two local minima: one at $\theta=\theta_m \simeq 41.4^o$ 
where $R_p\simeq 0.77$ and another at $\theta=\theta_R\simeq 43.7^o$ 
where $R_P\simeq 0.15$. Close to the local minimum at $\theta= \theta_m$ 
there is a local maximum at the (quasi-) total-refrection 
angle $\theta = \theta_0 \simeq 42.6^o$ where $R_p=0.98$. 
The very deep local minimum at $\theta_R$ is due to the 
surface plasmon resonance \cite{raether},  
while the behavior of the extrema at $\theta_m$ and $\theta_0$ 
is related to the thichness $d$ of the gold film: we have verified that 
when $d\to 0$ the reflectivity $R_p$  goes to zero 
at $\theta_m$ and it goes to one 
at $\theta_0$ (total reflection). These effects are clearly 
shown in Fig. \ref{fig2}, where we plot $R_p$ vs $\theta$ for 
different values of the thichness $d$ of the gold film. 

In the lower panel of Fig. \ref{fig1} we plot the phase 
$\phi_{\alpha}$ of the complex reflection 
coefficient $r_{\alpha}$ 
as a function of the incident angle $\theta$, using the same 
system parameters of the upper panel. The figure shows that, contrary 
to the phase $\phi_s$ of s-wave light (dashed line), 
the phase $\phi_p$ of p-wave light (solid line) 
changes abruptly near the resonant angle $\theta_R$. 

In Fig. \ref{fig2} we report the p-wave reflectivity $R_p$
as a function of the incident angle $\theta$ for 
four values of the thichness $d$ of the gold film. The figure shows that 
the resonant local minimum clearly appears only for $10$ nm $\lesssim 
d\lesssim$ $50$ nm. Indeed for both $d\gtrsim 50$ nm and $d\lesssim 10$ nm 
one finds that $R_p$ at $\theta_R$ is close to one. 

\begin{figure}
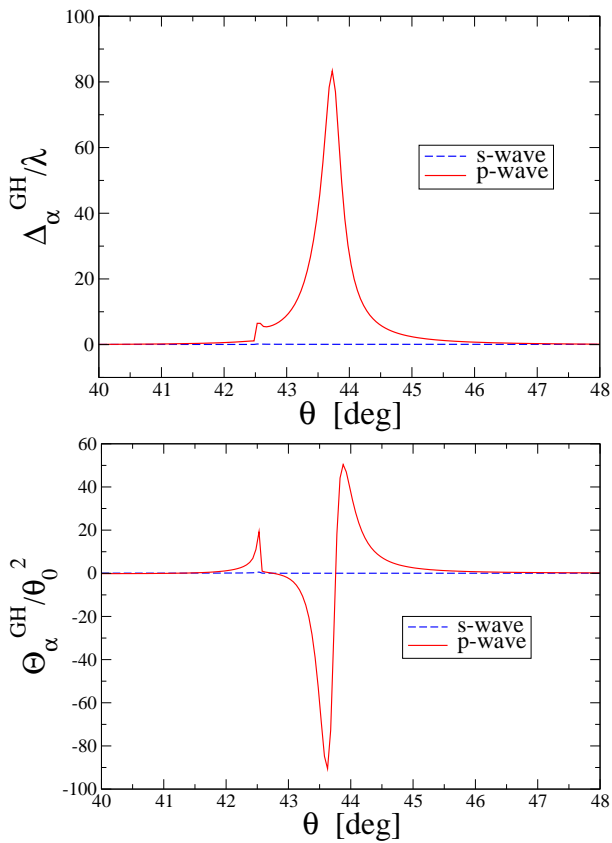

\centerline{\epsfig{file=spatial-gh.eps,width=8cm,clip=}}
\centerline{\epsfig{file=angular-gh.eps,width=8cm,clip=}}
\caption{(Color online). Upper panel: Goos-Hanchen spatial shift 
$\Delta_{\alpha}^{GH}$ as a function of the incident angle $\theta$
for s-wave ($\alpha=s$, dashed line) and 
p-wave ($\alpha=p$, solid line) monochromatic light. 
Lower panel: Goos-Hanchen angular shift $\Theta_{\alpha}^{GH}$
as a function of the incident angle $\theta$ 
for s-wave ($\alpha=s$, dashed line) and 
p-wave ($\alpha=p$, solid line) monochromatic light.
For both panels: 
Three-layer system with a gold film of thickness $d=30$ nm
at glass-air interface and light with wavelength $\lambda=830$ nm. 
$\theta_0$ is the angular spread of the incident beam.}
\label{fig3}
\end{figure}

The results shown in Figs. \ref{fig1} and \ref{fig2} 
are well-known \cite{raether} and 
confirmed by various experiments \cite{raether,guo}. Nevertheless, 
their consequences on GH and IF shifts are not yet fully explored. 
In particular, while the impact of the large derivative 
of the phase of the reflection 
coefficient for GH shift value has been previously studied
both numerically \cite{chen} and experimentally \cite{russi},  
the impact for IF shifts has not yet been explored. 

As discussed in detail by Aiello and coworkers \cite{aiello}, 
for a monochromatic beam of light with polarization $\alpha$ 
($\alpha=s,p$) and finite waist the total beam displacement 
$\delta_{\alpha}$ observed at distance $L$ from the reflection 
position is expressible as a linear combination 
of spatial shift $\Delta_{\alpha}$ and angular shift $\Theta_{\alpha}$, 
namely  
\beq 
\delta_{\alpha} = \Delta_{\alpha} + L \ \Theta_{\alpha} \; ,  
\eeq
under the condition $\Theta_{\alpha} \ll 1$. 
We have seen that when the shift $\delta_{\alpha}$ 
is parallel to the plane of incidence 
it is called Goos-Hanchen (GH) shift \cite{goos}, while when the 
shift $\delta_{\alpha}$ is normal of the plane of incidence it is called 
Imbert-Fedorov (IF) shift \cite{imbert}. 
Both spatial and angular shifts can be expressed in terms of the 
reflection coefficient $r_{\alpha}$, given by Eq. (\ref{hope}). 
In particular, 
in the case of linearly $\alpha$-polarized ($\alpha=s,p$) 
monochromatic light beam, with wavelength $\lambda$, incident 
angle $\theta$ and angular spread $\theta_0$, 
the shifts are given by \cite{aiello}
\beq
\Delta_{\alpha} = {\lambda \over 2\pi} \ Im[D_{\alpha}] \; , 
\quad\quad 
\Theta_{\alpha} = {\theta_0^2\over 2} Re[D_{\alpha}] \; , 
\label{essence}
\eeq
where 
\beq
D_{\alpha} = \left\{ \begin{array}{ll}
{d\over d\theta}\ln{\left( r_{\alpha}\right)}
& \mbox{in the GH case} 
\\
2i \cot(\theta) \ 
\left( {r_p + r_s \over r_{\alpha}}\right)     
& \mbox{in the IF case} \; . 
\end{array} \right.   
\label{essence1}
\eeq
Notice that, as explaned in Ref. \cite{aiello}, the IF shifts 
denote the spatial ($\Delta_{\alpha}^{IF}$)
and angular  ($\Theta_{\alpha}^{IF}$) separation between the two 
right-circularly and left-circularly polarized components of the 
reflected beam generated by the reflection-induced splitting 
of the $\alpha$-wave incident beam (see also \cite{li}). 
Eqs. (\ref{essence}) and (\ref{essence1}) show that while the 
spatial shifts depend on the presence of a 
imaginary component in the reflection coefficient, the 
angular shifts depend explicitly on the presence of a finite angular spread 
$\theta_0$ in the incident beam. 

\begin{figure}
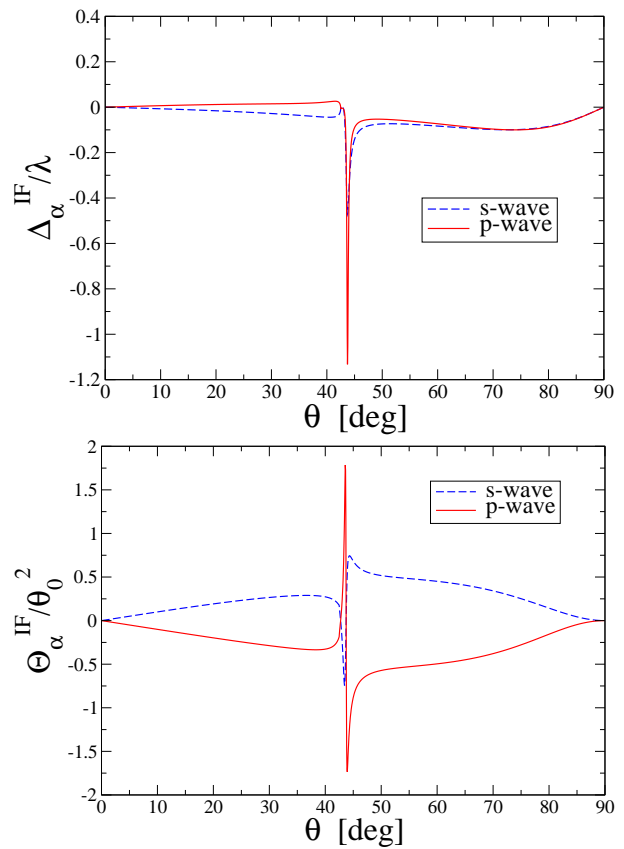

\centerline{\epsfig{file=spatial-if-all.eps,width=8cm,clip=}}
\centerline{\epsfig{file=angular-if-all.eps,width=8cm,clip=}}
\caption{(Color online). Upper panel: Imbert-Fedorov 
spatial shift $\Delta_{\alpha}^{IF}$
as a function of the incident angle $\theta$ 
for s-wave ($\alpha=s$, dashed line) and 
p-wave ($\alpha=p$, solid line) 
monochromatic light of wavelength $\lambda=830$ nm.
Lower panel: Imbert-Fedorov angular shift $\Theta^{IF}$
as a function of the incident angle $\theta$
for s-wave ($\alpha=s$, dashed line) and 
p-wave ($\alpha=p$, solid line) monochromatic light. 
For both panels: Three-layer system with a gold film of thickness $d=30$ nm
at glass-air interface and light with wavelength $\lambda=830$ nm. 
$\theta_0$ is the angular spread of the incident beam.}
\label{fig4}
\end{figure}

By using Eqs. (\ref{essence}) and (\ref{essence1}) 
with the reflection coefficients given by Eqs. (\ref{gf1}) and (\ref{gf2}) 
we calculate the four GH and IF shift of the light 
for our three-layer system. 

In the upper panel of Fig. \ref{fig3} we plot 
the Goos-Hanchen spatial shift $\Delta^{GH}$ 
as a function of the incident angle $\theta$ 
for s-wave (dashed line) and p-wave (solid line) monochromatic light. 
The s-wave light does not show any GH spatial shift. 
Instead for the p-wave light there is a extremely large 
(about 80 wavelengths) GH spatial shift in correspondence 
of the surface plasmon resonance (where $\theta=\theta_R\simeq 43.7^o$). 
Our results on the giant GH spatial shift at plasmon resonance 
angle $\theta_R$ are in full agreement with previous experimental 
data \cite{zhang}: also the presence 
of a small secondary peak at the total reflection angle 
$\theta_0\simeq 42.6^o$ is consistent with the p-wave experimental 
results of Yin, Hesselink, Lin, Fang and Zhang \cite{zhang}. 

Under the same system conditions (gold film of 
thickness $d=30$ nm and light with wavelength $\lambda=830$ nm), 
in the lower panel of Fig. \ref{fig3} we plot 
the Goos-Hanchen angular shift $\Theta^{GH}$ 
as a function of the incident angle $\theta$ 
for s-wave (dashed line) and p-wave (solid line) monochromatic light. 
This quantity, which has been recently measured by Merano, Aiello, 
van Exter and Woerdman \cite{exp-angular2} for a Gaussian laser beam 
at the air-glass interface, has not yet been observed 
in experiments with metals. 
The novel results shown in the lower panel of Fig. \ref{fig3} can be 
thus quite useful for next future experiments. The figure 
shows that $\Theta_s^{GH}$ is always zero. Instead, 
in analogy with the spatial 
GH shift $\Delta_P^{GH}$, the angular GH shift  $\Theta_P^{GH}$ 
is different from zero around the resonant angle $\theta_R$. 
Actually, $\Theta_P^{GH}$ is exactly 
zero at $\theta=\theta_R$ but it displays two positive peaks at 
$\theta_0$ and just above $\theta_R$,  and another very large 
negative peak just below $\theta_R$. 
There is a remarkable similarity between our results 
near the resonant angle $\theta_R$ and the experimental data 
obtained in Ref. \cite{exp-angular2} near the Brewster angle at the 
air-glass interface: in particular, both systems  
show a sudden change of the sign of $\Theta_p^{GH}$. 

In Fig. \ref{fig4} we report the IF shifts. In the upper 
panel of the figure we plot the IF spatial shift. 
This spatial shift is strongly enhanced at $\theta_R$ for both 
s-wave and p-wave light. However, this enhancement is much 
smaller than the GH one (in p-wave). 
In the lower panel of Fig. \ref{fig4} we show the IF angular 
shift. Both s-wave and p-wave angular shifts are nonzero 
for all incident angles (apart $\theta=0^o,90^o$) 
with a sudden change of sign around $\theta_R$. 
Notice that the IF angular shifts are much smaller with respect 
to the p-wave GH ones. Moreover, for s-waves the GH shifts are close 
to zero apart near the $\theta_0$ angle  
while the IF shifts are always nonzero and at resonance 
larger than the s-wave GH ones. This is due to the fact that  
IF shifts involve both $r_s$ and $r_s$ reflection coefficients. 

In conclusion, we have investigated Goos-Hanchen and Imbert-Fedorov spatial 
and angular shift for s-wave and p-wave light which reflects 
at the interface between glass and air with a thin film of gold. 
The presence of the metal induces a surface plasmon resonance 
which strongly suppress the reflectivity at a specific 
resonant angle of incidence and with a carefully chosen thickness. 
We have found that in correspondence 
of this critical angle both spatial and angular Goos-Hanchen 
shifts are remarkably enhanced in the case of p-polarized light. 
In addition, we have found a similar, but less pronounced, 
resonant effect on spatial and angular Imbert-Fedorov shifts 
for both s-wave and p-wave light. 
We stress that, up to now, only the spatial Goos-Hanchen shift 
has been experimentally observed \cite{zhang} 
for the three-layer system under consideration. 
For this reason, we think that our theoretical predictions can be 
a useful guide for next future experiments. 

The author thanks Michele Merano and Flavio Toigo 
for useful discussions and suggestions.

\end{document}